\documentclass[showpacs,amsmath,amssymb,twocolumn,aps,
        superscriptaddress]{revtex4-1}

 \usepackage{graphicx}
\usepackage{epsfig}              
 \usepackage{pstricks}
 \usepackage{pst-plot}
  \usepackage{psfrag}
 \usepackage{dcolumn}
 \usepackage{bm}

\usepackage{color}

\def\mn#1{}

\begin{document} 
\title{Transport in Graphene superimposed by 
a moving Electrical Superlattice Potential}

\author{J\"urgen Dietel}
\affiliation{Institut f\"ur Theoretische Physik,
Freie Universit\"at Berlin, Arnimallee 14, D-14195 Berlin, Germany}
\author{Hagen Kleinert}
\affiliation{Institut f\"ur Theoretische Physik,
Freie Universit\"at Berlin, Arnimallee 14, D-14195 Berlin, Germany}
\affiliation{ICRANeT, Piazzale della Repubblica 1, 10 -65122, Pescara, Italy}
\date{Received \today}

\begin{abstract}
We calculate dc-conductivities of ballistic graphene undulated by a   
overlying moving unidirectional electrical superlattice (SL) potential
whose SL-velocity is smaller than the electron velocity.  
We obtain no dependence of the conductivity on the 
velocity along the direction of the superlattice wavevector. In the orthogonal 
direction however,   
the dependence is strong on the velocity especially 
at voltages where a new Dirac point emerges for zero velocity. 
It is shown that the infinite 
graphene system can serve as 
an ideal motion detector at potentials where the first 
new Dirac point emerges. There  the conductivity 
is zero at vanishing  SL velocities and jumps to infinity when the SL starts 
moving.
For finite systems 
at voltages where the number of new Dirac points is of the order of the 
ratio of the electron velocity by the SL-velocity, the 
modifications to the conductivity of a moving SL is at least of similar 
magnitude as the conductivity of the stagnant SL.      
\end{abstract}


\maketitle

\section{Introduction} 
The electrical conductivity in suspended graphene samples 
show high mobilities where ballistic transport is seen for samples 
up to the micron length \cite{Morozov1,Du1, Bolotin1}.  
Due to the quasi-relativistic behavior of its electrons, 
graphene has a density of states  
proportional to the electronic energy which is zero at the neutrality point. 
As a consequence, this leads in ballistic graphene to the phenomenon
that the conductivity shows 
a universal finite behavior \cite{Castro1} whose precise value is 
still under debate \cite{Lewkowicz1}. It seems now that 
the universal conductivity in a wide range graphene sample 
with highly doped leads  
has the value $ \tilde{\sigma} = 4 e^2 /\pi h $ \cite{Katsnelson1, Tworzydlo1}, 
where in a system with vanishing small 
doped leads, it is $ \tilde{\sigma} = e^2 \pi/2 h $ 
\cite{Lewkowicz2}. 
Numerically, these two  values are quite close to each other. 
A small perturbation of the chemical potential of 
the graphene sample may be caused by applying an external gate voltage, 
this conductivity can change drastically, due the now finite density 
of states at the Fermi-energy. For an infinite 
large ballistic system, it becomes even infinite.  
Such an extreme sensitivity of the neutral graphene 
system on the  environment parameters  makes it attractive as a building block 
for nano-detectors. It was experimentally shown that  
graphene is a good  chemical sensor which is able to detect the 
dc-response changes due to  the adsorption of even single gas molecules on its  
surface \cite{Schedin1}. This high sensitivity is mainly due to the intrinsic 
low-noise properties of graphene. A more general review of possible 
graphene sensors can be found in Ref.~\onlinecite{Hill1}.     

Here we consider 
a ballistic graphene sample with an overlying slowly  
moving  unidirectional electrical SL. We 
calculate the longitudinal conductivities along and orthogonal to 
the SL wavevector as a response of a small external dc-field. 
This system  is considered as a possible model for a graphene-based 
nanomechanical motion detector. 

In the direction orthogonal to the wavevector of the SL
we obtain, especially at SL voltages where new Dirac points emerges in the 
non-moving SL, a high    
sensitivity of the conductivity values on the SL-motion. 
In the parallel direction  
our approximation produces no dependence  on the 
SL-velocity.

Graphene under the effect of a moving  SL can be 
realized for example by placing periodically patterned gate electrodes on either a moving  
underlying substrate or on a rested substrate 
where now the individual gate electrodes are activated appropriately with time
such that an effective moving SL is simulated.  
More directly, the experimental realization 
could be also carried out by using the coupling of the  graphene sheet 
to the electrical field of a surface acoustic wave 
on a piezoelectric substrate \cite{Thalmeier1} or 
to a charged moving membrane with ripples.  
   
It was recently shown explicitly for graphene that new Dirac points 
in the energy spectrum can be opened by imposing a non-moving 
SL on the graphene lattice \cite{Talyanskii1, Park1, Yankowitz1}. 
This leads to unusual conductivity properties in such systems 
\cite{Brey1, Park2, Dietel1, Burset1, Barbier2, Sun1}. 
These new Dirac points are accompanied with new energy valleys. 
Due to the technical complications in handling transport in 
a moving SL we will consider at first  
the transport contributions of the inner-valleys near the 
$ {\bf K} $ and $ {\bf K}' $ points in Sect.~III, then those 
of the outer-valleys in Sect.~IV. Note, that 
such a separation is not useful for the non-moving SL
as will be seen  in Sect.~IV. We start in Sect.~II 
by reconsidering first the 
lowest-band eigenvalues and eigenfunctions for the non-moving SL. 
 
We discuss  here the  most simple representation of a SL  
being a symmetric  two-step Kronig-Penney potential with 
a superlattice potential $ V(x)=V \chi(x) $ where 
 $ \chi(x)={\rm sg}[\sin(2 \pi x/d)]  $ (cf. Fig.~1). 
The function $ {\rm sg}[x] $  is the 
sign of $ x $, and $ d $ is 
the wavelength of the SL.  
In the continuum approximation,  
the graphene Hamiltonian under consideration near the Dirac point $ {\bf K} $   
is given by $ H_{v_s}=\hbar 
v_F (\sigma_x \partial_{x}/i + \sigma_y \partial_y/i) 
+ V(x+v_st) $ \cite{Castro1}. Here $ v_F $ is the Fermi velocity and 
$ \sigma_{x,y} $ are the Pauli matrices, while $ v_s $ is the  
velocity of the moving SL. Before starting, we mention here that we 
kept track of the most important in-line formulas in this paper in Tab.~1.    
This should enable the reader to better capture the structure of the paper.     Furthermore, we give a short guideline for reproducing the formulas used in 
this paper in App.~B.  

\section{Lowest band eigenvalues and eigenfunctions}

In the following, we solve the eigenvalue 
equation  $ H_{v_s} {\bf u}^{v_s}({\bf r}')=\epsilon \,  
{\bf u}^{v_s}({\bf r}') $ for a non-moving  SL ($v_s=0 $) 
by using the transfer matrix method \cite{Arovas1, Dietel1}.
For the energy dispersion  in the 
lowest band  we restrict ourselves to the lowest-lying oscillatory regime 
$   |\epsilon_s| d/\hbar v_F  \ll \tilde{V}, \tilde{\alpha}_{0}  $ 
and obtain \cite{Barbier1}  
 \begin{equation} 
\epsilon_s  = \! \!  s \hbar v_F  \tilde{\alpha}^2_0 
 \sqrt{k^2_x + 
|\Gamma|^2 k_y^2}      
   \,  ,         \label{20} \\
\end{equation}
with 
\begin{equation} 
 \alpha_{\epsilon_{s}}(x)  = \{[(\epsilon_s-V(x))/\hbar v_F]^2 -
      k_y^2\}^{1/2} d/2 \,.   \label{22} 
\end{equation} 
Here $ \Gamma=\sin[\alpha_{0}]e^{i \alpha_0}/\alpha_{0} $, 
$ \hat{\alpha}_0=\alpha_0/\tilde{V}$ where $ \tilde{V}= 
V  d/2 \hbar v_F $. The  
Bloch momentum in $ x $-direction is restricted to 
$ -\pi/d \le k_x \le \pi/d $. The parameter  
$ s=1 $ denotes the conduction band and $s=-1$ the valence band.  
 
The corresponding lowest-band eigenfunctions are  
$  {\bf u}^{v_s}(x,y) \! = \! e^{i k_y y} {\bf u}^{v_s}(x) $,  
in the fundamental zone $ 0 \le x \le d $ and  
for $v_s=0 $ reduces to 
\begin{equation} 
 {\bf u}^{0}_{s}(x)= \Lambda(x) {\bf u}^{0}_{s}(0)  \label{23}   
\end{equation} 
with 
\begin{equation} 
\Lambda(x)\!  = \! \lambda_0(x) \Theta \! \left(\! \frac{d}{2}-x \! \right) \! + \!  
\lambda_{d/2}(x) \lambda_{0}\! \left(\!\frac{d}{2}\!\right) \! 
\Theta\!\left(\! x- \frac{d}{2}\! \right) \,,  \label{24} 
\end{equation}  
where   
\begin{equation} 
\! \lambda_{x_0}(x)  =   \cos\left[\!   
\frac{\alpha_{\epsilon_{s}}(x) 2 (x-x_0)}{d} \! \right]  {\bf E}  +  
\frac{\sin \! \left[ \! 
\frac{\alpha_{\epsilon_{s}}(x) 2 
(x-x_0)}{d}\! \right] \!  }    {\alpha_{\epsilon_{s}}(x)}
  {\bf M} . \label{25} 
 \end{equation} 
Here $ {\bf E} $ is the unit matrix and 
\begin{equation} 
{\bf M}= k_y \sigma_{3} + i [\epsilon_s -V(x)] \sigma_2/\hbar v_F\,. 
\label{26}     
\end{equation} 
$ {\bf u}^{0}_{s}(0) $ is given in the oscillatory region $   
|\epsilon_s| d/\hbar v_F  \ll \tilde{V}, \hat{\alpha}_{0}  $ by 
\begin{equation} 
{\bf u}^{0}_{s}(0) \approx \frac{1}{N_u} \left(
\begin{array}{c} 
\frac{ \cos(\alpha_0) \sin(\alpha_0)}{\alpha_0} k_y d +i k_x d   \\
i \frac{1}{\hat{\alpha}_0^{2}}\, \frac{ \epsilon_s d}{\hbar v_F} + 
i \frac{\sin^{2}(\alpha_0)}{\alpha^2_0} \tilde{V}    k_y d       
\end{array} \right),  \label{28} 
\end{equation} 
where $ N_u $ in (\ref{28}) denotes  a normalization factor. 
From (\ref{20}) we obtain an oscillatory behavior of the lowest band 
eigenvalues as a function of $ k_y $  (cf. Fig.~1). New Dirac points emerge 
at $ {\bf k}=0 $ for $ \tilde{V} \in  \mathbb{N}  \pi $.
These are shifted along the y-axis in $ {\bf k}$-space 
for increasing $ \tilde{V} $.  
Note   
that the lowest band energy values beyond 
the oscillatory regime with  
momenta $ k_y^2 \gg (V/\hbar v_F)^2 $ 
scale like $ |\epsilon_s| \sim \hbar v_F |k_y| $ 
\cite{Arovas1, Dietel1}.

In the following, we discuss the transport 
contributions of electrons in the inner-energy valleys 
where $ k_y \ll V/\hbar v_F  $ and 
the outer-valleys where $ \hat{\alpha}_0 \ll 1 $ separately. 
Such a separation is possible for dc transport since as we will see 
in the following the dc electric field couples only electron states in the 
conduction and valence bands having the same Bloch momentum. 
Note that the Bloch momentum is conserved for a moving SL.  
The resulting time dependent state performs then a similar movement 
as is known under the Zitterbewegung in relativistic physics 
\cite{Lewkowicz2}.   
Taking into account all electrons in the valence band 
we obtain for large times an effective dc current.

\begin{table} 
 \begin{tabular}{|c|c|}
\hline 
$ \chi(x)={\rm sg}[\sin(2 \pi x/d)]  $ &  
$\xi(x)=\int^x_0 dx' \chi(x') $ 
\\[0.1cm] 
 $ \Gamma=\sin[\alpha_{0}]e^{i \alpha_0}/\alpha_{0} $  &  
 $ \hat{\alpha}_0=\alpha_0/\tilde{V}$  \\[0.1cm]
 $ \tilde{V}=  V  d/2 \hbar v_F $ & $ \tilde{k}_y= \hbar v_F k_y/V $  \\[0.1cm]
\hline\hline  
inner-valleys &  outer-valleys \\[0.1cm]
\hline 
$ \hat{\alpha}_0 \approx \frac{\pi n}{ \tilde{V}} $ &  
$ \hat{\alpha}_0= \frac{\pi n }{ \tilde{V}} $   \\[0.1cm] 
$ k^n_y d \approx 
2 [2 \tilde{V}(\tilde{V}- \pi n)]^{1/2}, k^0_y=0  $ &  
$ k^n_y d  =  2 [\tilde{V}^2 -(\pi n)^2]^{1/2} $\\[0.1cm]
$ \Gamma_n \approx 2 \left(1- \frac{ \pi n}{\tilde{V}}\right), 
\Gamma_0 = \frac{\sin(\tilde{V})}{\tilde{V}} $ & $ \Gamma_n = 
1-\left(\frac{\pi n}{\tilde{V}}\right)^{2} $ \\[0.1cm] 
 \hline
 \end{tabular}
\caption{Overview of the most important in-line formulas in this paper.} 
\end{table} 

\section{Inner-valley transport contributions} 

In the inner-valley regime $ k_y \ll V/\hbar v_F  $, 
the lowest-band eigenfunctions 
$ {\bf u}_s^{0}(x) $ (\ref{23})-(\ref{28}) for the non-moving system  
above are given by        
\begin{equation}
{\bf u}^{0}_{s}(x,t)=\!   \frac{1}{N_u}\!   \bigg[
\!   \binom{1}
 {1} \!\!  \left(\frac{-i k_x}{k_y \Gamma^*}    
\!  - \! \frac{i \epsilon_s}{\hbar v_F k_y \Gamma^*} \! \right) \! \phi_+(x) + \! \binom{-1}{1} \!   
\phi_-(x) \! 
\bigg]\! , \label{30} 
\end{equation} 
where $ \tilde{N}_u $ in (\ref{30}) denotes  a normalization factor.
$ \Gamma^* $ is the complex conjugate of $ \Gamma $.  
The phase factor $ \phi_{\pm } (x)$ is given by  
\begin{equation} 
 \phi_{\pm}(x,t)= 
\exp[i S^{v_s}_\pm(x,t)/\hbar]  \label{32} 
\end{equation} 
 with 
\begin{equation} 
 S^{0}_\pm(x,t) = \mp i \hbar 
\int_0^x dx' {\rm sg}\, [V(x')] \alpha_{\epsilon_s}(x')/(d/2) - 
i \epsilon_{s}  t                \label{34}    
\end{equation} 
 for $ v_s=0 $ where we extended (\ref{30}) by the last term in (\ref{34})  
chosen such that $ {\bf u}^{0}_{s} $ solves simultaneously the corresponding 
time-dependent Schr\"odinger equation (TSE). 
From (\ref{30}) we deduce the remarkable observation  that the inner-valley 
electrons do not backscatter at the potential steps. This phenomenon is 
well known for ordinary Dirac-fermions as Klein-paradox.

In the following, we use the    
inner-valley approximation 
\begin{equation}  
\alpha_{0}(x) \approx  \tilde{V} (1-\tilde{k}_y^2/2)  \label{36}  
\end{equation} 
 with 
$ \tilde{k}_y= \hbar v_F k_y/V $ in (\ref{20}), (\ref{30}) that is a
good approximation of the overall oscillatory behavior of the 
energy dispersion in Fig.~1. Similar approximations will also be used when 
solving the TSE for $ v_s\not =0 $ below.   
Finally we note that the missing of the $k_x,k_y $ dependence 
in the vector part $ (\mp 1,1)^T $ of both spinor components in (\ref{30}) is 
due to the inner-valley restriction 
$ k_y^2 \ll (V/\hbar v_F)^2 $. 

We obtain from (\ref{20}) that an entire set of    
$ 2 [\tilde{V}/\pi]+1 $ Dirac points exists  near $ {\bf K} $
where $ [x] $ is the lowest integer number smaller than $ x $. 
By using the inner-valley approximation (\ref{36}), 
these new Dirac points are located at 
 $ k^n_y d \approx 
2 [2 \tilde{V}(\tilde{V}- \pi n)]^{1/2} $ with $ n=1\ldots [\tilde{V}/\pi] $ 
and $ k^0_y=0 $ (restricting ourselves  to positive $ k_y $). 
The linearized energy spectrum 
around these Dirac points is given by $ \epsilon^n_s  = \! \!  s \hbar v_F  
  [k_x^2 + \Gamma_n^2 (k_y-k^n_y)^2]^{1/2} $ where the effective $y$-velocity 
coefficient is given by $ \Gamma_n \approx 2 (1- n \pi /\tilde{V}) $ for 
$ n=1,\ldots,[\tilde{V}/\pi] $, and $ \Gamma_0 = \sin(\tilde{V})/\tilde{V} $ 
for the central valley. The magnitude $ \hat{\alpha}_0 $ for  
$ k_y=k^n_y$ is given by $ \pi n/ \tilde{V} $.    
Below, we shall also need the $k_y $-momentum 
spacings between the right and left-energy crest and the Dirac point. 
The spacing for the right crest is given by 
$\Delta  k_y^{n,R} d \approx \pi 
   [\tilde{V}/2(\tilde{V}-\pi n)]^{1/2} $ for $ n=1,\ldots, 
[\tilde{V}/\pi] $ and 
$ \Delta k_y^{n,L}= \Delta k_y^{n,R} $ for the left crest positions where 
$ n=1 \ldots [\tilde{V}/\pi]-1 $. For the central crest distance  we obtain 
$ \Delta k_y^{0,R}d = \Delta k_y^{0,L} d \approx 
(k_y^{[\tilde{V}/\pi]}d )^3/[(k_y^{[\tilde{V}/\pi]} d )^2+ 4\Gamma_0 \tilde{V}^2] $ 
and $ \Delta k_y^{[\tilde{V}/\pi],L} =k_y^{[\tilde{V}/\pi]}-\Delta k_y^{0,L} $. 
Finally we note that the inner-valley formula with 
$ \tilde{k}_y^2 \ll 1 $  considered in this subsection 
is valid for the valleys  
$ 1- \pi n/\tilde{V} \ll 1 $ with $ n \not= 0 $ and also the central valley 
$ n=0 $.

In the following 
we solve the TSE  
$ i \hbar \partial_t {\bf u}_s^{v_s}(x,t)=H_{v_s}{\bf u}_s^{v_s}(x,t) $ 
with the initial condition $ {\bf u}^{v_s}_s(x,0)={\bf u}^{0}_{s}(x) $  
for $t=0 $ in the oscillatory regime by using the  above approximations. 
Note that by using the characteristic-method we can solve the TSE  
without approximation for $ k_y =0 $. 
This leads again to (\ref{30}) 
where now $ S^{v_s}_{\pm}  $ is $ v_s $-dependent. 
Instead of doing this  
explicitly, we can generalize this procedure to any non-zero   
 $ k_y^2 \ll (V/\hbar v_F)^2 $ by the    
Hamilton-Jacobi Ansatz
\begin{equation} 
\! -\! \frac{\partial S^{v_s}_\pm}{\partial t} \! = \! - 
\hbar v_F {\rm sg}[V(x+v_st)]  
\sqrt{ (\partial_x S^{v_s}_\pm)^2 \! + \! k_y^2} + V(x+v_s t)  \label{40}
\end{equation}    
with the boundary condition that $ S^{v_s}_\pm(x,0)=S^{0}_\pm(x,0) $. 
Due to the local uniformity    
of $ V(x+v_s t) $ in position and time we obtain local uniform solutions   
of (\ref{40}). 
That this approach leads to a TSE solution in 
the oscillatory regime is due to the fact that the general solution 
can be written as 
\begin{equation} 
 {\bf u}^{v_s}_s(x,t)
  \! \approx \! \!  \sum_{s,k_x} \! \! a_{s,k_x} \!  \binom{ \! s \frac{k_x}{|k_x|}\! }{1}  
 e^{-\frac{i}{\hbar} t [\hbar v_F s 
\sqrt{k_x^2+k_y^2}+V(x+v_st )]} e^{ik_x x}    \label{50}
\end{equation}  
in the inner-valley regime $ k_y^2 \ll k_x^2 $. The complex variables 
$ a_{s,k_x} $ are local uniform functions in the $ (x,t) $-plane. 
We will show below that $ a_{s,k_x} $  is non-zero for only two special 
$ k_x $-values which moreover fulfil the inner-valley regime 
condition $ k_y^2 \ll k_x^2 $. 

\begin{figure}
\begin{center}
\includegraphics[clip,height=5cm,width=8.5cm]{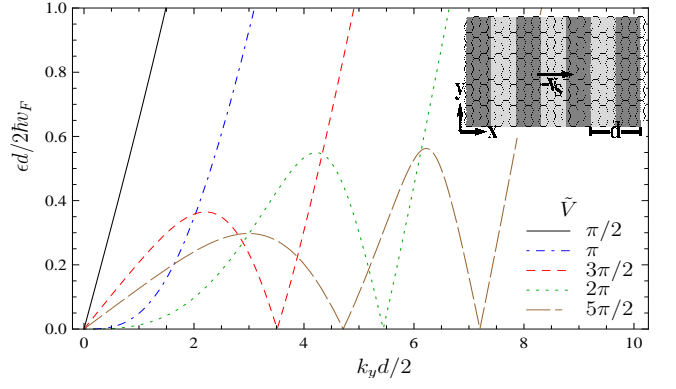}\\[0.08cm]
\caption{(Color online) Lowest Bloch band energy spectrum 
for $ k_x=0 $ as a function of dimensionless momenta $ k_y $ for 
various SL potential strengths  $ \tilde{V} $ (the full lowest band  
energy spectrum can obtained by using its mirror symmetry 
with respect to the x,y-axis).
Here we used the transfer matrix method \cite{Arovas1, Dietel1}. Inset 
shows a graphene layer with an overlying moving SL in x-direction.      
}
\end{center}
\end{figure} 

We now solve  
(\ref{40}) by using a generalized characteristic  
method for the Hamilton-Jacobi equation  
that is well known in the semi-classical approach to quantum mechanics 
\cite{Maslov1}. This is based on the one-particle mechanical 
trajectory of a relativistic particle and anti-particle in a step-potential. 
The calculation is outlined in App.~A.

After some manipulation we obtain  the result
\begin{equation} 
S^{v_s}_\pm(x,t) \approx  S^{v_s,x}_\pm + S^{v_s,t}_\pm \label{52} 
\end{equation} 
with
\begin{align} 
&  \frac{S^{v_s,t}_\pm}{\hbar}= -t \frac{\epsilon_s}{\hbar} \,, \label{55} \\ 
& \frac{S^{v_s,x}_\pm}{\hbar}  \! \!   
= \! \! \pm [A_\pm \xi(x^* \! + \! v^*_s t) \! \! + \! \! B_\pm 
\xi(x^* \! \mp \! v^*_Ft) 
\! \! + \! \! C_\pm t \chi(x^* \! \mp \! v^*_Ft)]\,.  \nonumber 
\end{align}
Further we have $\xi(x)=\int^x_0 dx' \chi(x') $, $ x^*=x-(v^*_s-v_s)t $  and 
\begin{align}
& A_\pm \!= \! \mp  \!  \frac{V }{\hbar(v_s \pm v_F)} \! \left[ 
1 \! - \!   \tilde{k}_y^2  \frac{v_s \pm \frac{1}{2} v_F}{
v_s \pm v_F} \! + \! \tilde{k}_y^2 Z^2_{\pm}  \right],    
 \nonumber  \\ 
& B_\pm \! = \! -  \frac{V}{\hbar (v_s \pm v_F)} \left[ \frac{v_s}{v_F} - 
\frac{\tilde{k}_y^2 v_s }{2v_F }  \frac{v_s}{v_s\pm v_F} \mp  
\tilde{k}_y^2 Z^2_{\pm}\right] ,  \nonumber
\\
& C_\pm= \pm \frac{V}{\hbar} \tilde{k}_y^2 Z^1_{\pm} \,, \label{60} \\ 
& v^*_F \! = \! v_F \!\left[\! 1- \! \frac{\tilde{k}_y^2 }{2}\! 
\frac{(v_s^2+v_F^2)^2}{(v_s^2-v_F^2)^2} \! \right]   
\! , \! v^*_s \! = \! v_s\! \left[\! 1- \! \tilde{k}_y^2 \! 
\frac{v_F^2 (v_s^2+v_F^2)}{(v_s^2-v_F^2)^2}
\! \right]  \nonumber 
\end{align}  
where $ Z^1_{\pm} =  v_s^3  (v_s \pm v_F)/(v_s^2- v_F^2)^2 $ and  
$ Z^2_{\pm}= v_s^2 /(v_s^2- v_F^2) $.   
We restrict here the solution of (\ref{40}) to small 
velocities $ v_s \lesssim  v_F $.  
  
Next we calculate the dc-response in the moving SL system. 
This is done in the gauge  
$ {\bf A}=- c {\bf E} (t-t_0)  \Theta(t-t_0) $ 
assuming $ t_0 \le  0 $ in general. Since 
$ \tilde{\sigma}_{ii}(t) $ does not depend on $ t_0 $ for $ t \gg 0 $ 
we set immediately $ t_0 = 0$.  
The total Hamiltonian in the continuum approximation 
is then given by 
$ H_{A} = H_{v_s}+ \hbar v_F (e/c) (\sigma_{x} A_x+\sigma_{y} A_y) $. 
The corresponding TSE-solution which we expand to first 
order in $ {\bf A} $ and assume 
it to satisfy the initial condition 
$ {\bf u}_{A}(t=0)={\bf u}^0_{s} $ is denoted by $ {\bf u}_{A} $.  
From this solution we obtain   
the conductivity in the $ i$-th direction by   
$ \tilde{\sigma}_{ii} = \lim_{E \to 0} e 
v_F (\langle  {\bf u}_{A}(t) \sigma_{i} 
{\bf u}_{A}(t) \rangle / E)$                                       where 
$ {\bf A}= -c E {\bf e}_i  t $. Here  $ {\bf e}_i $ is the unit vector 
in the $i$-th direction. 
The conductivity in the $i$-th direction in the lowest energy level 
approximation valid for $ t \to \infty $ and $ v_s \lesssim  
v_F, V d /\hbar $  is then given by \cite{Lewkowicz2} 
\begin{equation} 
   \tilde{\sigma}_{ii}(t)\! = \! \frac{-4 e v_F}{(2 \pi)^2}
 \int_{\rm BZ}  \! \! \! d^2k  \mbox{Re}[\langle {\bf u}^{v_s}_{-1}(t)|  
\sigma_{i}| {\bf u}^{v_s}_{+1}(t) \rangle \xi_+(t) ]    \label{75}     
\end{equation} 
with 
\begin{align}   
& \quad \xi_+(t)=  i \frac{ e v_F}{\hbar}  
 \int_{0}^t dt' \, t' {\cal T}^{v_s}(t') \label{80}       \\ 
& \qquad \quad \, \, = \! i \frac{e v_F}{\hbar} 
  \left( \! \! t \! \! \int_{t''=-\infty}^t  \! \! \! \! 
\! \! \!\! \! \!  dt'' \! \! - \! \! \int_{t'=0}^t \! \! \! \! \! dt' \! \! 
\int_{t''=-\infty}^{t'} \! \! \! \! \! \! \! \! dt'' \! \right) 
{\cal T}^{v_s}(t'') 
 \nonumber 
\end{align} 
and the transition matrix element $ {\cal T}^{v_s}(t)=
\langle {\bf u}^{v_s}_1(t)| \sigma_{i} |{\bf u}^{v_s}_{-1}(t) \rangle$. 
By inserting (\ref{80}) in (\ref{75})  
the term proportional to $ t $ cancels  in an improved 
tight-binding approximation since it can be written as 
$ t \int_{\rm BZ}  d^2k 
\, \partial_{k_i}  
\langle {\bf u}^{v_s}_{-1}(t)| J_i |{\bf u}^{v_s}_{-1}(t) \rangle  =0 $, 
where $ J_i $ is the tight-binding current operator for $ {\bf A}=0 $ 
\cite{Lewkowicz2}. 
Here we used the fact that the exact tight-binding wave functions 
are smooth at the Brillouin zone boundary.
Summing the Fourier series  $ 
\sum_{\omega_n} {\cal \hat{T}}^{v_s}_r (\omega_n) e^{i \omega_n t} \equiv 
e^{-i t \Delta \epsilon / \hbar }{\cal T}^{v_s} (t) $
where $ \Delta \epsilon = \epsilon_{1}-\epsilon_{-1} $ we obtain
for large times   
\begin{align} 
& \xi_+(t)= i \frac{e v_F}{\hbar}  \sum_{\omega_n} 
\bigg\{  \frac{{\cal \hat{T}}^{v_s}_r(\omega_n) 
}{ (\omega_n +\Delta \epsilon/\hbar  - i \delta)^2 }\left(e^{i (\frac{\Delta \epsilon}{\hbar}+ \omega_n)t } -1\right)
  \nonumber \\
&  
\qquad \qquad - i t  \,   
\frac{[ {\cal \hat{T}}^{v_s}_r(\omega_n)- {\cal \hat{T}}^{0}_r(\omega_n) ]}{\omega_n +\Delta \epsilon/\hbar - i \delta}  \bigg\} \,. \label{90}  
\end{align}  
Here $ \delta $ is an infinitesimal positive number. 

In the following we calculate the contribution of 
every energy valley to the 
momentum integral in (\ref{75}) separately, i.e.
\begin{equation}  
 \tilde{\sigma}_{ii}(t) = \sum_{n=0..[\tilde{V}/\pi]} \tilde{\sigma}^n_{ii}(t)
(2- \delta_{n,0}) .   \label{93} 
\end{equation} 
For large times one can restrict  
the $ k_y$-integrals of Eq.~(\ref{75}) 
to the neighbourhood of the valley center $k_y^n $ 
setting immediately  $ k_y \approx k_y^n $ in 
$ {\cal \hat{T}}_r^{v_s} $. This leads then with (\ref{30}), (\ref{75})
and (\ref{90})  
to the following momentum integrals during the calculation  
of $ \tilde{\sigma}^n_{ii} $                                      
\begin{align}
& \int\limits_{\rm n-th\,    valley}  \!\!\!\! d^2k 
\frac{(2 \hbar v_F k_x)^2}{(\Delta \epsilon)^2} 
e^{-\frac{i}{\hbar} \Delta \epsilon t }  
 \xi_+(t) \label{100} \\
&  \qquad = \frac{i e}{2 \hbar} \frac{1}{\Gamma_n} \bigg\{
\! \! \int_{-\infty}^{+\infty} \! \! \! \! \! \!\! \! \!  dk_x \! \! \! 
\int\limits_{-\Delta k_y^{n,L}\Gamma_n}^
{\Delta k_y^{n,R} \Gamma_n}\! \! \! \! \! \! \! \! dk_y  \frac{k_x^2}{k^3}  
F^{v_s}(k)     \nonumber \\
& \qquad \qquad  \; \; - \int_0^{2 \pi} \! \! \! \!  d \vartheta  
k_\vartheta \sin^2(\vartheta) [F^{v_s}(k_\vartheta)-F^{0}(k_\vartheta)] \bigg\}  \nonumber 
\end{align}  
with 
\begin{align} 
\quad  F^{v_s}(k) = 
\frac{{\cal \hat{T}}^{v_s}_r(\omega_n) e^{i \omega_n t }}{\omega_n + 2v_F 
k  -i \delta} 
(1- e^{-i  (2 v_F k +\omega_n)t})     \label{105} 
\end{align}
and $ k_\vartheta = 
(\tan^2(\vartheta)+1)^{1/2} \Gamma_n
\{\Delta k_y^{n,R} \Theta[\cos(\vartheta)]+ \Delta k_y^{n,L}
\Theta[-\cos(\vartheta)]\}$ where $ \Theta(x) $ is the Heaviside function.
The right-hand side of Eq.~(\ref{100}) was calculated by the help of a partial 
integration.

In the calculation of $ \tilde{\sigma}_{ii} $ via (\ref{75}), the quantities
\begin{eqnarray}   
  {\cal P} & \equiv &  \frac{1}{d} 
\int^d_0 dx \exp^{i (S^{v_s,x}_+- S^{v_s,x}_-)} \,, \nonumber \\  
{\cal C}_m & \equiv & \sum_{\omega_n \approx 2 m \pi   v^*_F/d}   
\hat{|{\cal P}|}(\omega_n) (2-\delta_{n,0}) e^{i \omega_n t}  \label{110}    
\end{eqnarray}  
are relevant where $ \hat{|{\cal P}|} (\omega_n) $ are the 
Fourier components  of $ |{\cal P}|(t) $. More precisely, 
$ {\cal C}_m $ with $ m>0 $ are the positive components  
for frequencies $ 2 \pi (m-1/2) v^*_F/d \le 
\omega_n \le 2 \pi (m+1/2) v^*_F/d  $ under the restriction that 
$ \omega_n\ge 0$ for $ m=0 $. 
A straight-forward calculation leads with (\ref{55}) for $ v_s \lesssim 
 v_F $ to 
\begin{align} 
& \mbox{Re}[{\cal C}_m]= 
\sum_{\sigma \in \{\pm \}} \cos(C_- t) 
 B_- X(m ,m,m,\sigma)  \label{120} \\
& \!  - \! \sin(C_- t) \left[ \!  \sigma \frac{2 \pi m}{ d}+
\dot{\xi}(2 v^*_s t) B_+  \! 
\right]\! 
  X(m,m,m+1,\sigma) \! +\!   {\rm Ex}, \nonumber  \\ 
& \mbox{Im}[{\cal C}_m]= 
 \sum_{\sigma \in \{\pm \}}  \sigma \cos(C_-t) B_- X(m,m,m+1,\sigma) \nonumber \\ 
&  +\sigma  \sin(C_- t) \left[ 
\sigma \frac{2 \pi m }{d} +\dot{\xi}(2 v^*_s t) B_+ 
  \right]
   X(m,m,m,\sigma) \! + \! {\rm Ex} \nonumber 
\end{align}
where 
\begin{align} 
& 
 X(n_{\omega},n_1,n_2,\sigma) \approx  - \frac{1}{A d^2} 
\frac{8   (2- \delta_{n_{\omega},0})}
{B^2_- - [\dot{\xi}(2 v^*_s t) B_+ +  
\sigma  \frac{2 \pi n_{\omega}}{d} ]^2}  \nonumber \\
& \sin \bigg[\! \! A \frac{d}{4} \! + \!                
  n_1 \frac{ \pi }{2}\bigg] 
 \sin\bigg[\! \!   - \! B \frac{d}{4}\!  + \! B_+ \xi(2 v^*_s t)- 
C_+t \chi(2 v^*_s t) \nonumber \\
&  -\sigma n_{\omega} \frac{2 \pi v_F^* t}{d}  
 \! \left(\! 1-\frac{v^*_s}{v^*_F}\! \right)  
 +n_2  \frac{\pi } {2} \bigg] \,. 
  \label{122}  
\end{align} 
The term Ex    in (\ref{120})    
stands for the foregoing expressions with    
interchanged $ B_+ \Leftrightarrow B_- $, $ C_+ \Leftrightarrow -C_- $ 
and switched  sign of $ v_s^*$. 
We used further the abbreviation $ A \equiv A_{+}+ A_{-}$, $ B\equiv 
 B_++B_- $.  
 
We are now able to calculate the conductivity contribution  
$ \tilde{\sigma}^n_{ii} $ of the $n$-th energy valley by using (\ref{75}), 
(\ref{100}), (\ref{120}), leading to  
\begin{align}
& \tilde{\sigma}^n_{xx} \approx  \frac{e^2}{h} \frac{\pi}{2} \hat{\alpha}_0^2 
\frac{1}{\Gamma_n}  
\,, \label{125} \\ 
& \tilde{\sigma}^n_{yy} \approx   \frac{e^2}{h}\frac{\pi}{2} 
   \frac{1}{\hat{\alpha}_0^2} 
\frac{1}{\Gamma_n} 
 \sum_{\sigma \in \{ L,R \}} 
\mbox{Re}[{\cal C}_0+{\cal C}_1] 
 \bigg\{\mbox{Re}[{\cal C}_0]
  \nonumber  \\
& +
 \mbox{Re}[{\cal C}_1] \left[\frac{1}{\pi}
\{\vartheta^n_\sigma -\frac{1}{2} 
\sin(2 \vartheta^n_\sigma)\} - I_2 
 \left(\frac{\Gamma_n \Delta k_y^{n,\sigma} d}{\pi} ,\vartheta^n_\sigma\right) \right] \nonumber  \\
&  - 
\mbox{Im}[{\cal C}_1]\left[
I_1\left(\frac{\Gamma_n \Delta k_y^{n,\sigma} d}{\pi}
\right) +I_3
\left(\frac{\Gamma_n \Delta k_y^{n,\sigma} d}{\pi}\right) \right] 
\bigg\} \,.  \nonumber 
\end{align}   
Terms containing $ {\cal C}_m $ with $ m \ge 2 $ are neglected here which 
can be justified numerically.     
The angle $ \vartheta^n_\sigma $ is given by $ \tan(\vartheta^n_\sigma)= 
(\Gamma_n d \Delta k_y^{n,\sigma})/
[\pi ^2-(\Gamma_n d \Delta k_y^{n,\sigma})^2]^{1/2} $.          
The functions  $ I_1,\ldots,I_3 $ are calculated from (\ref{100}) for 
$ t \to \infty $ as   
\begin{align} 
& I_1(x)= \frac{2}{\pi^2}  x +\frac{2}{\pi^2}   \, {\rm sgn}   [1-x^2]\mbox{Re}\bigg\{
\frac{a(x)}{4} \bigg[
\csc^2\left(\frac{a(x)}{2} \right)   \nonumber \\ 
& \! \! -  \! \sec^2 \! \left(\! \! \frac{a(x)}{2} \! \! \right) \! \! \bigg] 
\! \! \!     
 - \! \! \! \! \! \sum_{\sigma \in \{ \pm \}} \! \! \!\!  \! \sigma    \! \left[   
a(x) \! \log(1 \! + \! \sigma e^{i a(x) })   
\! + \! i {\rm Li}_2 (\sigma 
e^{i a(x)})\right]\! \! \bigg\},      \nonumber \\   
& I_2(x,\varphi)= \frac{2}{\pi x^2} \sin^3(\varphi) |\cos(\varphi)|\,,   \label{140} \\         
& I_3(x)=\frac{4}{\pi^2} x \left(\sqrt{x^2-1} \, a(x) -1 \right),    \nonumber 
\end{align}   
where  $ a(x) \equiv \arctan(1/\sqrt{x^2-1}) $ and $ {\rm Li}_2 $ is the 
dilogarithm function.

Here the term $ I_1 $ is calculated from the first summand in the 
integral on the right hand side of (\ref{100}). For the calculation of 
$ I_2 $, $I_3 $ we used the last term in (\ref{100}). Furthermore we took 
into account  in Eq.~(\ref{125}) the degeneracy of the 
$ {\bf K} $ and $ {\bf K}'$ valleys and the spin degeneracy.    
       
We obtain from (\ref{125}) that the conductivity $ \tilde{\sigma}^n_{xx} $ 
does not depend on $ v_s $, whereas 
$ \tilde{\sigma}^n_{yy} $ shows a strong $ v_s $-dependence.  
Eq.~(\ref{110}) shows that for $ v_s = 0 $ the only finite term in 
$ \tilde{\sigma}^n_{yy} $ is the term proportional 
$ \mbox{Re}[{\cal C}_0]^2 $ in (\ref{125}). In order to derive this term 
we made use of the following integral 
$ \lim_{t \to \infty} \int^\infty_0 dk \sin(2 k t)/k = \pi/2 $. With the help of 
$ \mbox{Re}[{\cal C}_0]=|\Gamma| $ for $ v_s =0 $,     
$ \tilde{\sigma}^n_{yy}$ is reduced to 
$ \tilde{\sigma}^n_{yy}= \delta_{n,0} \Gamma_n e^2\pi /2 h  $.
Furthermore we find for $ \tilde{\sigma}^0_{xx} $ a divergence 
at SL potentials where $ \tilde{V} \in \mathbb{N} \pi $ for general velocities.
The same thing holds for $ \tilde{\sigma}^0_{yy} $ but here 
we must demand $ v_s >0 $ where now $ \mbox{Re}[{\cal C}_0] \not= 
|\Gamma| $, i.e. $ \mbox{Re}[{\cal C}_0] \not= 0 $ in general.  
The origin of these divergences comes from the  
vanishing of $ \Gamma_0$ in the denominator in the right hand side of  
(\ref{125}). This term is already existent in (\ref{100}). 
The reason of this  vanishing is based on the flatness 
of the energy band (\ref{20}) at the central 
Dirac-point in the $ k_y $-direction at SL potentials 
where $ \tilde{V} \in \mathbb{N} \pi $. In the next section  
(cf. Eq.~(\ref{320})) we show for the $ v_s=0 $ conductivity, 
by going beyond the inner-valley approximation 
used here, that $ \tilde{\sigma}_{yy} $ is exactly 
vanishing only for $ \tilde{V}= \pi $. All this leads us to the following     
remarkable fact:

  An infinite large SL graphene 
sample is an {\it ideal motion detector} 
at SL potentials where the first new Dirac point emerges, i.e. 
at $ \tilde{V}=\pi  $.    
 There $ \tilde{\sigma}_{yy} $ is vanishing for $ v_s=0 $ 
and jumps to infinity for $ v_s \not= 0 $.     

From (\ref{90}), (\ref{100}), the divergence of 
$ \tilde{\sigma}^0_{yy} $ 
at $ \tilde{V} \in \mathbb{N} \pi $ has its origin in the approximation that  
we used an infinite ballistic time $ t \sim t_b $ in calculating the response.
This is not really valid for a  finite system where $ t_b \sim L/v_f $ and 
$  L $ is the length of the sample.   
By repeating the discussion below (\ref{80}) but now use the energy (\ref{20})  
$ \epsilon_s \approx   s \hbar v_F (k_x^2+ d^4 k_y^6/64 \tilde{V}^4)^{1/2} $  
at small momenta for $ \tilde{V}  \in  \mathbb{N} \pi $ leads to  
$ \tilde{\sigma}^0_{ii} $ in (\ref{125}) with a finite cut-off 
at $ 1/\Gamma_0 \sim  (\tilde{V} t_b)^{2/3} $. In the following we 
calculate  from (\ref{90}) the conductivities $ \sigma^0_{ii} $ 
at  $ \tilde{V} \in 
 \mathbb{N} \pi $ 
in leading order in $ 1/t_b $ for $ t_b \to \infty $. 
The results are  
$ \tilde{\sigma}_{ii} \approx \tilde{\sigma}^0_{ii} $ with 
\begin{align}
&  \tilde{\sigma}^0_{xx} \! = \!  \frac{e^2}{h} \frac{\sqrt{3}}{4^{2/3} 
\sqrt{\pi}}  
\frac{\Gamma(1/6)\Gamma(1/3)}{\Gamma(2/3)}
\! \! \left(\! \! \tilde{V} \frac{v_F t_b}{d}\! \! \right)^{2/3}\,,  \label{150} \\ 
&  \tilde{\sigma}^0_{yy} \! = \!  \frac{e^2}{h} \frac{\sqrt{3}}{4^{2/3} \sqrt{\pi}}  
\frac{\Gamma(1/6)\Gamma(1/3)}{\Gamma(2/3)}
\! \! \left(\! \! \tilde{V} \frac{v_F t_b}{d}\! \! \right)^{2/3} \! \! \!   \! \! \! \! \! 
\! \mbox{Re}[{\cal C}_0+{\cal C}_1] 
\mbox{Re}[{\cal C}_{00}]  \nonumber   
\end{align}     
where $ {\cal C}_{00} = (v_s/d) \int^{d/v_s}_0 dt \, {\cal C}_{0} $. 
By using (\ref{120}) we obtain  
\begin{align} 
& {\rm Re}[{\cal C}_{00}]= \frac{32 \sin\left(A \frac{d}{4} \right)}{ B_+ B_- A d^3} 
\bigg[ \! \!  \cos\left(\! \! \{B_+ \! - \! B_-\}\frac{d}{4}\right)-\cos\left(B \frac{d}{4} \right) 
\! \! \bigg]. \label{170}
\end{align}      

\begin{figure*}
\begin{center}
\includegraphics[clip,height=6.5cm,width=12.5cm]{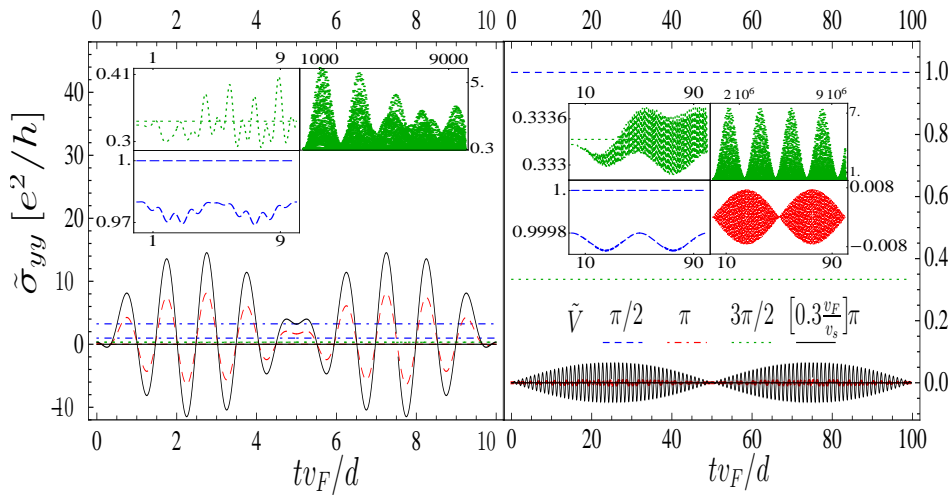}
\caption{(Color online) We show the conductivity $ \tilde{\sigma}_{yy} $ 
orthogonal to the SL calculated   
for velocities $ v_s/v_F=0.1 $ (left panel) 
and $ v_s/v_F=0.01 $ (right panel) as a function of the dimensionless time.  
We plot the curves for various SL-potentials $ \tilde{V} $ 
by using (\ref{125}) for $ \tilde{V} \notin \mathbb{N} \pi $,    
(\ref{150}) for $ \tilde{V} \in \mathbb{N} \pi $ and ballistic times  
$ (v_F t_b/d)^{2/3}=1000 $. The horizontal curves show $ \tilde{\sigma}_{yy}$
for $ v_s=0 $. Insets in both panels  
show a zoom in of the corresponding curves in the main panels 
(upper left: $ \tilde{V}=3 \pi/2 $, lower left: 
$ \tilde{V}=\pi/2 $, lower right: $ \tilde{V}=\pi$). 
We also add for $ \tilde{V}=3 \pi/2 $ in the upper right 
insets $ \tilde{\sigma}_{yy} $ for larger times.}     
\end{center}
\end{figure*}
 
In Fig.~2 we plot $ \tilde{\sigma}_{yy} $ for $ v_s/v_F=0.1 $ (left panel) and 
$ v_s/v_F=0.01 $ (right panel), as well as for 
$ v_s=0 $ (horizontal curves) 
at various $ \tilde{V} $-values. 
The most interesting $ \tilde{V} $-values 
are where for a certain $ v_s $ the signal $ \tilde{\sigma}_{yy} $ is largest. 
In particular, the signal to background 
ratio, i.e., $ \tilde{\sigma}_{yy} $  divided by 
$ \tilde{\sigma}_{yy} $ for $ v_s=0 $, should be large.
We obtain from the figure and (\ref{125}) as well 
as (\ref{150}) that for a finite system and $ v_s \ll v_F $,  
the SL potential region where $ \tilde{V} \sim v_F/v_s $ and 
$ \tilde{V} \in  \mathbb{N} \pi $ gives the best results.  
We plot this in Fig.~2 for $ \tilde{V}= [0.3 v_F/v_s] \pi $. This is chosen 
so that the curves do not show a higher-order 
$ v_s $-Fourier behavior according to (\ref{120}).      
We note that in principle 
a graphene velocity detector based on a SL considered here 
could also attain a large signal to background conductivity for small 
velocity differences by using large SL potentials $ \tilde{V} \sim 
v_F/\Delta v_s $.  This is due to the phase factors in (\ref{120}). 

Beside the oscillation frequencies $ \sim 2 \pi v^*_F/d $ and  
$ \sim 2 \pi v^*_s/d $ we also find from (\ref{120}) and Fig.~2 
a much smaller oscillation frequency $ \sim C_\pm $ for the conductivity 
contribution of the side-valleys 
becoming relevant only on very large time scales. 
One can show that due to its non-zero velocity, the SL    
transfers additional energy and momentum to an electron 
passing its potential steps such that the electron  velocity oscillates between 
$ \pm v_F(1-\tilde{k}_y^2/2) $ and $ \pm v_F[1-\tilde{k}_y^2
(v_F \pm v_s)^2/2 (v_F \mp v_s)^2] $. Due to this velocity difference 
the electron picks up an additional oscillating phase  
proportional to $ t $ represented by the last term in (\ref{55}). This 
leads to the long wave-conductivity oscillations shown in Fig.~2.

To complete our discussion, we finally  
 calculate the quasi-particle velocities in the 
$ x $ and $y $-direction   for electrons in the 
$ {\bf u}_{\pm 1}^{v_s} $ state where now 
$ E_{\rm dc}=0 $. The knowledge of these velocities is useful in quantum 
pumping experiments \cite{Thouless1, Prada1}.  
We obtain from (\ref{20}) 
and (\ref{30}) 
\begin{align} 
& v_x= v_F \langle {\bf u}_{\pm 1}^{v_s}|\sigma_{xx} |{\bf u}_{\pm 1}^{v_s}\rangle
 = \frac{\partial \, \epsilon_{\pm}}{\hbar \partial k_x} \,, \label{200} \\
& 
  v_y= v_F 
 \langle {\bf u}_{\pm 1}^{v_s}|\sigma_{yy} |{\bf u}_{\pm 1}^{v_s} |\rangle = 
 \frac{\partial \, \epsilon_{\pm}}{\hbar \partial k_y} |{\cal P}| \approx 
 \frac{\partial \, \epsilon_{\pm}}{\hbar \partial k_y} 
\mbox{Re}[{\cal C}_0+{\cal C}_1].          \nonumber 
\end{align}    
This means that similar to the above conductivity considerations  
we obtain no time dependence of $ v_x $, in contrast to 
$ v_y $. As in the     
non-moving  system \cite{Park3} 
there is a  collimination of the electron motion in 
$ x$-direction, i.e., $|v_{y}| \ll  |v_{x}| $ 
for potentials were  $ \tilde{V} \approx  \mathbb{N} \pi $ and momenta 
$ k_y $ near the central Dirac point. Here we use that 
$ |{\cal P}| \le 1 $.     

\section{Outer-valley transport contributions} 
Next, we discuss the conductivity contributions of the 
outer-energy valleys where 
$ \hat{\alpha}_0 \ll 1 $. We obtain from (\ref{20}) 
that the new Dirac points are located at 
$ k^n_y d  =  2 [\tilde{V}^2 -(\pi n)^2]^{1/2} $ where the linearized energy 
spectrum around these points is given by   
$ \epsilon_s  = \! \!  s \hbar v_F  
  [\hat{\alpha}_0^{4} k_x^2 + \Gamma_n^2 (k_y-k^n_y)^2]^{1/2} $. The  
effective $y$-velocity 
coefficient is now given by $ \Gamma_n = 
[\tilde{V}^{2}-(\pi n)^{2}]/\tilde{V}^{2} $ and  
$ \hat{\alpha}_0= \pi n / \tilde{V} $. This means that the 
outer-valley regime $ \hat{\alpha}_0 \ll 1 $ is fulfilled for those valleys 
where $ \pi n / \tilde{V} \ll 1$.

We obtain now from Sect.~II for the space evolution operator (\ref{25}) 
of the non-moving system
\begin{equation}  
 \lambda_{x_0}(x) \approx \frac{1}{\alpha_{\epsilon_s}(x)} 
{\bf M} \sin\left[2 \alpha_{\epsilon_s}(x)  \frac{(x-x_0)}{d}\right]   \label{240} 
\end{equation} 
 to leading order 
in $  \hat{\alpha}_0  $. The corresponding lowest-band eigenfunctions 
$ {\bf u}^0_{s} $ can be interpreted by electrons which 
are fully backscattered close to the potential steps for 
$ |\epsilon_s| d/\hbar v_F  \ll \hat{\alpha}_{0} $. 
This is just the opposite situation of the inner-valley
transport contributions discussed in Sect.~III where we got a complete 
transmission through the potential steps.  This interpretation is 
even justified by discussing the scattering of electrons   
on a single potential step in the momentum regime $ \hat{\alpha}_0 \ll 1 $.
In this regime $ {\bf u}^{v_s}_s $ can now be written as in (\ref{50}) 
with the substitution of the spinor part  
$ (s k_x/|k_x|,1) \rightarrow (- i s k_y/|k_y|,1) $. For the moving lattice 
we concentrate ourselves in the following on a 
particle moving in a potential $ \pm V $ in the region 
$ -v_st  \le x \le d/2-v_st $.    

We now determine a complete set of functions $ v_\pm^{j}(x,t) $ 
fulfilling the quasi-relativistic Klein-Gordon equation with   
a potential $ V(x)= \pm V $  in the region 
$ -v_s t \le x \le -v_s t+d^*/2 $. They further  satisfy    
the zero-boundary conditions $ v_\pm^{j}(-v_s t,t)=
v_\pm^{j}(-v_s t+d^*/2,t)= 0$. These properties unambiguously 
define the functions $ v_\pm^{j}(x,t) $. 
The distance $ d^* $ has a small modification 
to the distance $ d $ for 
$ |\epsilon_s| d/\hbar v_F  \ll \hat{\alpha}_{0} $ determined by  
$ \alpha_{\epsilon_s} d^*/d= \pi n $ for the $ n$-th energy valley, 
i.e. $ \alpha_{0}= \pi n $. The wavefunctions $ v_\pm^{j}(x,t) $ 
consist of a superposition of two Klein-Gordon wave-function solutions.
The momenta of both Klein-Gordon wave-functions can be formally 
derived from the zero-boundary conditions.  
More concrete the 
two corresponding momenta are given by a particle initial momentum 
and its reflected momentum at the boundary. 
In the quasi-nonrelativistic limit 
valid for $ v_F |k_x^j/k_y|, |v_s|  \ll v_F $ we obtain for these momenta 
$ k^j_x \pm  v_s |k_y|/v_F $ and $ -k^j_x \pm  v_s |k_y|/v_F $ with  
$ j \in \mathbb{N} $ and $ k_x^j= 2 \pi j /d^* $ in the potential 
$ V(x) = \pm V $ .  
The restriction on the quasi-nonrelativistic limit is justified 
for the outer-valley transport contributions in the case 
$ v_s \ll v_F $.
This leads to   
\begin{align} 
&  v_\pm^{j}(x,t)= \frac{2}{\sqrt{d^*}}  e^{\pm i (\hbar v_F \sqrt{k_x^j+k_y^2} 
- V) t/ \hbar  } e^{\mp i (1/2) (v^2_s/v_F) |k_y| t} \nonumber \\ 
&\qquad \qquad \times e^{\pm i (v_s |k_y|/v_F) (x+v_s t) }
 \sin[k^j_x (x+v_s t)] \,. 
\label{250} 
\end{align} 
By using (\ref{240}) with (\ref{23})-(\ref{28}),     
the wavefunction $ {\bf u}^{v_s}_s $ is then given by
\begin{align} 
& {\bf u}^{v_s}_s (x,t)=
-i \frac{d}{2 \alpha_0} \bigg\{ {\rm sg}[k_y V(x+v_s t)] 
\frac{ \cos(\alpha_0) \sin(\alpha_0)}{\alpha_0} k^2_y d  \nonumber \\
&+ 2 {\rm sg}[k_y] 
\frac{\epsilon_s \tilde{V} }{\hbar v_F \hat{\alpha}_0^{2}}
+ i {\rm sg}[V(x+v_s t)] |k_y| k_x d \bigg\}   \label{260}    \\   
& 
\times \left({\rm sg}[k_y V(x+v_s t )] i \atop 1 \right) \sum_j 
c^{\rm sg[V(x+v_s t)]}_j  \, 
v^j_{\rm sg[V(x+v_s t)]} (x,t) \nonumber 
\end{align} 
with  
\begin{equation}   
c^{\rm sg[V(x+v_s t)]}_j =  \int^{d^*/2}_0  \! \! \! \! \! \! \! dx \;    
(v_{\rm sg[V(x+v_s t)]}^{j})^*(x,0)  \sin\left(\alpha_{\epsilon_s} 
\frac{2 x}{d} \right). \label{270} 
\end{equation} 

With this  wavefunction in hand we are now prepared 
to calculate the conductivities $ \tilde{\sigma}^n_{ii} $ 
for the outer-valleys $ \pi n / \tilde{V} \ll 1$. By using 
(\ref{75}) with (\ref{90}) and (\ref{260}) we obtain for 
the conductivities 
\begin{equation} 
 \tilde{\sigma}^n_{xx} \approx 0 \quad  , \quad   
\tilde{\sigma}^n_{yy} \approx \frac{e^2}{h} \frac{\pi}{2}
\frac{1}{\hat{\alpha}^2_0} \Gamma_n Y(v_s)   \label{280}     
\end{equation}    
with $ Y(v_s) $ is given by 
\begin{equation} 
Y(v_s)= \sum_{i,j>0} \frac{n^2}{i^2} |c_i|^2 |c_j|^2 \left(\delta_{i,j} +
2 \delta_{i >j}\right) \,. \label{290} 
\end{equation}
Here we use $ |c_i| =|c^{\pm}_i| $ 
and $ c_i, c_j $ in (\ref{290}) and (\ref{270}) are calculated with 
$ d^* \rightarrow d $.   

We obtain from (\ref{280}) that the transport contributions of the 
outer-valleys corresponding to  $ \hat{\alpha}_0\ll 1 $ show 
no time-fluctuations. 
This is not based on the quasi-nonrelativistic approximation 
used above.   
We show in Fig.~3 $ \tilde{\sigma}^n_{yy} $ for the outer-valleys 
and various SL potentials $ \tilde{V} $ and velocity fractions 
$ v_s/v_F  $. Most pronounced, the curves on the right panel show 
a conductivity peak at valley indices where 
$ n \approx n_0 $. Here $ n_0 $ is given by  
$ n_0= [|v_s / v_F| (\tilde{V}^2 -(\pi n)^2)^{1/2}/\pi] $. 
This conductivity peak is also 
observed from (\ref{280}) and (\ref{290}) by taking into account that 
in a rough approximation we have  $ |c_i|^2 \approx ( \delta_{i,n+n_0}+
\delta_{i,|n-n_0|})/2 $ leading to 
\begin{equation} 
 Y(v_s) \approx \frac{1}{4} \frac{n^2}{|n-n_0|^2}+\frac{3}{4} 
\frac{n^2}{|n+n_0|^2}  \,. \label{295}      
\end{equation}  
All this means that for $ \tilde{V} \gtrsim  \pi v_F/v_s $ with 
$ \tilde{V} \gg 1 $ we obtain a large conductivity signal 
where the conductivity modification due to the motion of the SL is of 
similar magnitude as the conductivity value of the non-moving SL. 
Something similar applies for the detection of small velocity 
differences $ \Delta v_s $ where now we have 
$ \tilde{V} \gtrsim  \pi v_F/\Delta v_s $
in order to obtain a large signal to background value.  
By comparing the conductivity values $ \tilde{\sigma}^n_{yy} $ for the 
inner-valleys (\ref{125}), Fig.~2  and the outer-valleys  (\ref{280}), Fig.~3 
we obtain at least for $ \tilde{V} \gg 1 $ and 
$ \tilde{V} \not\approx \mathbb{N} \pi $ that the outer-valley contributions 
are dominant. 

\begin{figure}
\begin{center}
\includegraphics[clip,height=5.5cm,width=8.4cm]{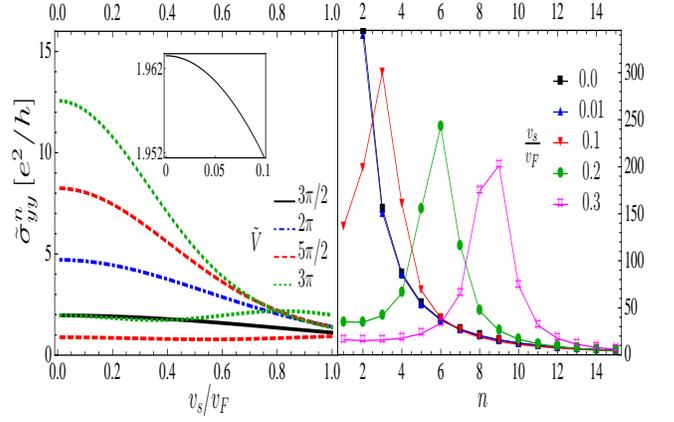}
\caption{(Color online)  Left panel: 
Outer valley-conductivities $ \tilde{\sigma}^n_{yy} $ (\ref{280}) 
of the $n$-th electron side-valley 
as a function of $ v_s/v_F $ for certain SL potentials. For 
$ \tilde{V}=5 \pi/2 $ and $ \tilde{V} = 3 \pi $, which both consists 
of two side-valleys, the upper curve corresponds to 
the valley index $ n=1 $ and the lower curve to $ n=2 $. 
Inset shows a zoom in of $ \tilde{\sigma}^1_{yy} $ for $ \tilde{V}=3 \pi/2 $. 
Right panel: $ \tilde{\sigma}^n_{yy} $ for $ \tilde{V}= 30 \pi $ 
as a function of the valley index 
$ n $ for certain SL velocities $ v_s/v_F $.     
}     
\end{center}
\end{figure}

Next, we calculate the effective particle velocities for 
electrons in the outer-valley defined in (\ref{200}), where now again 
$ E_{\rm dc}=0 $.   
By using (\ref{250}) and (\ref{260}) we obtain  
\begin{equation} 
 v_x= \frac{\partial \, \epsilon_{\pm}}{\hbar \partial k_x} \quad , \quad    
  v_y=\frac{\partial \, \epsilon_{\pm}}{\hbar \partial k_y}.   \label{300}        \end{equation}    
This shows that there is no $ v_s $-correction term in contrast to 
the inner-valley case (\ref{200}) for $ v_y $. 
This is caused by the fact that in the 
outer-valley regime electrons are approximately 
fully reflected, and thus the total 
probability of finding an electron between $ -v_s t $ and $ -v_s t+d/2 $ is 
conserved.

The non-trivial dependence of the conductivities on the SL-velocity 
forced  us to treat the 
conductivity contributions for the inner and outer-valleys separately. 
This separation is no longer necessary when calculating the conductivities
for the non-moving SL. For this we use the full oscillatory wave-function 
 (\ref{23})-(\ref{28}) with (\ref{75}), (\ref{100}). This leads us to the 
following $ v_s=0 $-conductivities
\begin{equation} 
  \tilde{\sigma}^n_{xx} =  \frac{e^2}{h} \frac{\pi}{2}
\hat{\alpha}^2_0   \frac{1}{\Gamma_n}  
\quad, \quad  \tilde{\sigma}^n_{yy} = \frac{e^2}{h} \frac{\pi}{2}
\frac{1}{\hat{\alpha}^2_0} \Gamma_n\,.    \label{320}     
\end{equation}    
Note here that the magnitudes of $ \hat{\alpha}_0 $ and $ \Gamma_n $  
correspond to the outer-valley values discussed above Eq.~(\ref{250}) 
for $ n \not= 0 $ and to the $n=0 $ values discussed above 
Eq.~(\ref{40}). Similar expressions as in (\ref{320}) were calculated 
before within the dc vector potential gauge 
$ {\bf A}= 0 $, leading as in pristine graphene  
to a small overall numerical prefactor correction to our result (\ref{320})   
\cite{Burset1}. The disadvantage of the calculation in 
Ref.~\onlinecite{Burset1} lies in the strong dependence of this 
prefactor on the order of taking 
the zero-temperature, zero-frequency, and zero-damping limit. 
This does not happen in our calculation \cite{Lewkowicz2}.

\section{Summary} 
Summarizing, we have considered the dc-transport in neutral graphene 
undulated by 
a unidirectional moving superlattice potential with 
$ v_s \lesssim  v_F, V d/\hbar $. 
While the 
response along the direction of the 
SL wave-vector is vanishing, the dependence is dramatic 
in the orthogonal  direction. 
In particular we find for potentials where the first new Dirac 
point emerges, i.e.,  at $ \tilde{V}= \pi $, 
that the infinite large 
graphene sample is a perfect motion detector. The orthogonal 
dc-conductivity is vanishing for zero velocity and jumps to infinity at 
non-zero SL-velocity. A large conductivity signal 
with a high signal to background ratio is 
reached for the finite but large 
graphene system when $ \tilde{V} \in \mathbb{N} \pi $.  
The time fluctuating contribution to the 
conductivity is largest when $ \tilde{V} \sim v_F/v_s $.
All this was derived from the inner-valley contributions 
to the conductivities. 

Next we have calculated the conductivity contributions of the outer-valleys. 
The conductivity contributions parallel to the SL-wavevector are vanishing.
In the orthogonal direction they are large, time-independent and exhibit  
a peak as a function of the valley index. 
For $ \tilde{V} \gg 1 $ and $ \tilde{V} 
\gtrsim \pi v_F/v_s   $ the conductivity modifications due to a 
moving SL are of similar magnitude as the conductivity 
values of the stagnant SL. Note that for $ \tilde{V} \gg 1 $ 
the outer-valley conductivity 
contributions are dominate over the 
inner-valley contributions, at least 
for $ \tilde{V} \not\approx \mathbb{N} \pi $. 
Finally, we have calculated the conductivities of the non-moving SL 
without the need of a separate calculation for the inner and 
outer-valleys.     
 
Due to its intrinsic  low-noise level 
\cite{Schedin1}, our results could be useful for graphene as a 
nanophysical motion detector device, or even for general sensors 
based on the surface acoustic wave technology \cite{Mamishev1}.

\begin{appendix}

\section{Solving the Hamilton-Jacobi equation Eq.~(\ref{40})}
Here we outline the calculation of (\ref{55}) by solving the 
Hamilton-Jacobi equation (\ref{40}) to first order 
in $ \tilde{k}^2_y $. This is done with  the help of a generalized 
characteristic method \cite{Maslov1}. The solution is based 
on the one-particle quasi-relativistic orbit  $ x(t)$ 
in a moving potential $ V(x+v_st)  $. With the help of this solution, 
$ S_{\pm}^{v_s}(x,t) $ is given by the action integral  
\begin{align} 
& S_{\pm}^{v_s}(x,t)=\int_{0}^{t}d t' \bigg\{ 
\frac{v_F \hbar^2 k^2_y}{\sqrt{p^2(t')+\hbar^2 k_y^2}} \; 
{\rm sg}[V(x(t')+v_st')] 
\nonumber \\ 
& \qquad \quad \; \; \, -
V(x(t')+v_st')   \bigg\}
+S_{\pm}^0(x_0,0) \,.          \label{a10}
\end{align} 
Here $ x(t') $ is the particle trajectory with $ x(0)=x_0 $, $ x(t)=x $.
The particle-momentum  is given by
\begin{equation}  
p(t')= \partial_x S_\pm^{v_s}(x(t'),t')  \label{a15}  
\end{equation} 
and 
the quasi-relativistic velocity by 
\begin{equation}    
 \dot{x}(t')=  - v_F  {\rm sg}[V(x(t')+v_st')] 
\frac{p(t')}{\sqrt{p^2(t')+\hbar^2 k_y^2}} \,.    \label{a20}
\end{equation} 
We note now that it is much easier to 
determine $ x(t') $ by solving the set of equations above for small 
$ \tilde{k}_y $, instead 
of solving the second-order quasi-relativistic Newton equation. 
From this we obtain (\ref{55}). 

\section{A guideline to reproduce the formulas}
Here we give a short guideline for readers 
who would like to reproduce the formulas in this paper.       
 
 \subsection{ Eqs.~(\ref{52})-(\ref{60})} 
We first solve (\ref{a10})-(\ref{20}) in leading order in $ \tilde{k}_y^2 $, 
i.e. for $ k_y=0 $. This leads with 
(\ref{a20}) to the particle velocities up to 
the next leading order in $ \tilde{k}_y^2 $. We obtain  
\begin{eqnarray} 
\dot{x}(t) & = & v_0 \,  \delta_{{\rm sg}[V(x_0)],{\rm sg}[V(x(t) +v_s t)]} 
 \label{a30}   \\
& & + v_1
(1-\delta_{{\rm sg}[V(x_0)],{\rm sg}[V(x(t) +v_s t)]}) \nonumber
\end{eqnarray} 
with $ v_0=\pm v_F(1-\tilde{k}_y^2/2) $ and 
$ v_1= \pm v_F[1-\tilde{k}_y^2(v_F \pm v_s)^2/2 (v_F \mp v_s)^2] $.   
With these velocities in hand one can derive the particle's action 
$ S_{\pm}^{v_s} $ to order $ \tilde{k}^2_y $ 
by using (\ref{a10}), (\ref{a20}). Here we have used
the idendity  
$ x_0= x- (v_0+v_1)t /2 + \Delta x_0 $ where 
\begin{eqnarray} 
\Delta x_0 & \approx  &  - {\rm sg}[V(x_0)] \frac{v_F^2 v_s}{v_s \mp v_F } 
\frac{1}{v_s^2- v_F^2} \nonumber \\
& & \times [\xi(x+v_s t)- \xi\left(x-(v_0+v_1)t/2  
\right)]  \label{a35} 
\end{eqnarray}
during the derivation.
Eq.~(\ref{a35}) is valid in the next to leading order in $ \tilde{k}_y^2 $.
It connects the starting point $ x_0 $ of the trajectory with its 
end point  $ x $.  

We calculated  $v_0 $, $v_1 $ in (\ref{a30}) by using the  
approximation $ |\epsilon_s| \ll V $. Going beyond this approximation  
could lead for $ S_{\pm}^{v_s,t} $ (\ref{55})  
to small possible additional terms of the order 
$ \pm t \epsilon_s \tilde{k}_y^2 (v_s/v_F) $. Such terms would then result   
in a small time-independent numerical prefactor correction in the oscillatory  
side-valley conductivity $ \tilde{\sigma}^n_{yy} $ for $ n>0 $ of the order 
$ (k^n_y d/\tilde{V})^2  (v_s/v_F)^2 $ (\ref{125}). The conductivities 
$ \tilde{\sigma}^n_{xx} $ would get a similar small prefactor correction.  

Finally we note, that by setting  $ Z^1_\pm, Z^2_\pm=0 $ 
in (\ref{52})-(\ref{60}),  the corresponding action 
$ S^{v_s}_{\pm} $  is given by (\ref{a10}) 
where now the particle trajectory and the particle momentum 
is calculated from the uniform velocity  
$ v_1= v_0= \pm v_F(1-\tilde{k}_y^2/2) $.

\subsection{Eqs.~(\ref{110})-(\ref{122})} 
In order to derive (\ref{120}), (\ref{122}) from (\ref{110}) we 
used that $ |A| \gtrsim |B| $ for $ v_s \lesssim v_F $. Then we  obtain  
for not too large ballistic times $ C_\pm t_b \ll 1 $ but also for 
large times $ C_\pm t_b \gtrsim 1 $ where now we have to restrict ourselves  
to the most relevant low-frequency Fourier components 
${\cal C}_m $ with $ m \lesssim 1 $, that   
\begin{align}  
& |P|(t',t) \approx  \frac{4}{A d} 
\sin\! \! \bigg[\! \! (A+B) \frac{d}{4} \! - \! 
B_+ \xi (-v^*_F t'-v^*_s t) \nonumber \\
&\qquad \qquad   -B_-\xi (v^*_F t'-v^*_s t) -\! C_+\!  t \chi(-v^*_F t'-v^*_s t)
                        \nonumber \\ 
&\qquad \qquad  -C_-t 
\chi(v^*_F t' - v^*_s t) \bigg]\,,    
 \label{a40}
\end{align}
where $ |P|(t)=|P|(t,t) $.   
In order to calculate $ {\cal C}_m $
(\ref{110}) we can use for $ v_s \lesssim v_F $ the approximation 
$ C_m \approx 
( d / v_F^*) \int^{v_F^*/d}_0 dt' |P|(t',t) e^{- i 2 \pi m v_F^* t' /d} $,   
which then leads to the expressions (\ref{120}), (\ref{122}).

\subsection{Eqs.~(\ref{125})-(\ref{140})} 
The integrals $ I_1(x) $ and $ I_3(x) $ which are the terms proportional 
to $ {\rm Im}[{\cal C}_1] $ at the right bottom of 
Eq.~(\ref{125}), are calculated by using (\ref{100}) and (\ref{105}) 
\begin{align} 
&  I_1(x)= \frac{1}{2 \pi^2} \! \! 
\int_{-x}^{+x}\! \!  \! \! dk_y \! \! \int_{- \infty}^{+\infty} 
\! \! \! \! \! \! \! \! dk_x \! \!  
\frac{k_x^2}{\sqrt{k_x^2+ k_y^2}^3} \! \! \sum_{\sigma \in \{\pm\}} \! \!
\frac{\sigma}{ \sqrt{k_x^2+ k_y^2}-\sigma},   \nonumber \\
&  I_3(x)= \frac{1}{2 \pi^2} \! \! \int^{2 \pi}_0 \! \! \! \! d \vartheta \! \! 
 \sum_{\sigma \in \{\pm\}}
\sigma \frac{\sin^2(\vartheta) \sqrt{1 + 
\tan^2(\vartheta)}}{\sqrt{1 + 
\tan^2(\vartheta)} - \sigma /x} \,. 
\end{align}   
The terms proportional to $ {\rm Re}[{\cal C}_1] $ in  Eq.~(\ref{125})  
were also derived from (\ref{100}) with (\ref{105}) by making 
use of the identity $ \lim_{t \to \infty} \frac{1}{f(k)} \sin[ f(k)t ]
= \pi \delta(f(k)) $ for an arbritary function $f $. 
Here $ \delta(x)$ is the Dirac delta-function.             

\subsection{Eqs.~(\ref{270})-(\ref{290})} 
Here we use (\ref{260}) with (\ref{250}) in (\ref{90}) and (\ref{75}). 
With the help of a small $ \epsilon_s $ expansion of the exponents in 
(\ref{250}) we obtain (\ref{270})-(\ref{290}) by using that 
$ \lim_{t \to \infty} \int^\infty_0 dk \sin(2 k t)/k = \pi/2 $. 
Note that we get a contribution only from the first 
term in Eq.~(\ref{90})  in this calculation 
which leads to the final result (\ref{280}) 
for $ t \to \infty $.   
\end{appendix}

\end{document}